\documentclass[12pt]{article}

\usepackage{amsmath}
\usepackage{graphicx}
\usepackage{booktabs}
\usepackage{amsfonts}
\usepackage{hyperref}
\usepackage{amssymb}
\usepackage{caption}
\usepackage{subcaption}
\usepackage{color}

\renewcommand{\theequation}
{\arabic{section}.\arabic{equation}}

\makeatletter
\def\eqnarray{ \stepcounter{equation} \let\@currentlabel=\theequation
 \global\@eqnswtrue
 \global\@eqcnt\z@
 \tabskip\@centering
 \let\\=\@eqncr
 $$\halign to \displaywidth\bgroup\@eqnsel\hskip\@centering
 $\displaystyle\tabskip\z@{##}$&\global\@eqcnt\@ne
 \hfil$\displaystyle{{}##{}}$\hfil
 &\global\@eqcnt\tw@$\displaystyle\tabskip\z@{##}$\hfil
 \tabskip\@centering&\llap{##}\tabskip\z@\cr}
\makeatother

\makeatletter
\def\@arrayacol{\edef\@preamble{\@preamble \hskip .5\arraycolsep}}
\def\array{\let\@acol\@arrayacol \let\@classz\@arrayclassz
\let\@classiv\@arrayclassiv \let\\\@arraycr\def\@halignto{}\@tabarray}
\makeatother

\makeatletter
\newcounter{subeqncnt}
\def\thesubeqncnt{\alph{subeqncnt}}
\def\subequations{\begingroup%
   \stepcounter{equation}\edef\@tempa{\theequation}%
   \let\c@equation\c@subeqncnt\c@subeqncnt\z@
   \edef\theequation{\@tempa\noexpand\thesubeqncnt}}

\makeatother

\newcommand{\be}{\begin{equation}}
\newcommand{\ee}{\end{equation}}

\newcommand{\bea}{\begin{eqnarray}}
\newcommand{\eea}{\end{eqnarray}}
\newcommand{\nn}{\nonumber}

\def\CM {{\cal M}}

\def\CZ {{\cal Z}}

\def\Det{{\rm Det}}

\begin{document}

\setlength{\baselineskip}{7mm}
\begin{titlepage}
 \begin{flushright}
{\tt NRCPS-HE-74-2017}
\end{flushright}

\begin{center}
{\Large ~\\{\it   Classical Limit Theorems   
\vspace{1cm}

and 
\vspace{1cm}

High Entropy MIXMAX random number generator

}

}

\vspace{1cm}

{\sl Hayk Poghosyan${}^{a,b}$, Konstantin Savvidy${}^{a}$ and
 George Savvidy${}^{a}$

\bigskip
${}^{a}$ \sl   Demokritos National Research Center, Ag. Paraskevi,  Athens, Greece\\
${}^{b}$ \sl Yerevan Physics Institute, 2 Alikhanyan Br., 375036, Yerevan,  Armenia \\

\bigskip

}
 
\end{center}
\vspace{30pt}

\centerline{{\bf Abstract}}

We investigate the interrelation between  the distribution 
of stochastic fluctuations of independent random variables in probability 
theory  and the distribution of time averages in deterministic  Anosov C-systems.  
On the one hand, in probability theory,  our interest dwells on three basic topics: the laws of large numbers, the central limit theorem  and the law of the iterated logarithm for sequences of real-valued random variables. On the other hand we have chaotic, uniformly hyperbolic   Anosov  C-systems defined on tori which have mixing of all orders and nonzero Kolmogorov entropy. These extraordinary  ergodic  properties of  deterministic Anosov C-systems ensure that the above classical limit theorems for sums of independent random variables in probability theory are  fulfilled by the time averages for the sequences generated by the C-systems.  
The MIXMAX generator of pseudorandom numbers represents the  
 C-system  for which the classical limit theorems are  fulfilled.

\noindent

\end{titlepage}

\pagestyle{plain}

\section{\it Introduction}
 Our intention in this article is to consider the behaviour of deterministic 
Anosov C-systems in parallel with the classical limit theorems of probability theory,  demonstrating that they possess the properties which are 
inherent to the independent and identically distributed random variables defined in probability theory.

We investigate the interrelation between  the distribution 
of stochastic fluctuations of independent random variables in probability 
theory  and the distribution of time averages in deterministic  Anosov C-systems.  
On the one hand, in probability theory,  our interest dwells on three basic topics: the laws of large numbers, the central limit theorem  and the law of the iterated logarithm for sequences of real-valued random variables \cite{kolmogorovprob,khinchine,kolmo2,hartman,gendenkokolmogorov,prochorov,liapounoff,liapounoff1,berry,esseen,esseen1,petrov} 
. On the other hand we have chaotic, uniformly hyperbolic  Anosov  C-systems defined on tori which have mixing of all orders and nonzero Kolmogorov entropy \cite{anosov,kolmo,kolmo1,sinai3,rokhlin,rokhlin2,sinai4,gines,bowen0,ruelle,krilov,Savvidy:2017mew,Savvidy:2015ida}. These extraordinary  ergodic  properties of  Anosov C-systems ensure that the above classical limit theorems for sums of independent random variables in probability theory are  fulfilled by the time averages for the sequences generated by the C-systems \cite{leonov,chernov3,rozanov,rather}.  
The MIXMAX generator of pseudorandom numbers represents the homogeneous 
 C-system  for which the classical limit theorems are  fulfilled \cite{Savvidy:2017mew,Savvidy:2015ida,yer1986a,konstantin,Savvidy:2015jva}.

The present paper is organised as follows. In section two we shall overview the 
 classical limit theorems in probability theory. In section three the basic properties of  the Anosov C-systems will be defined, their spectral properties and the entropy
 will be presented.  In section four a parallel between the classical limit theorems of probability theory  and behaviour of deterministic dynamical C-systems will be derived 
 and the mapping dictionary between the two systems will be established.  
We shall analyse the 
 law of large numbers,  the central limit theorem  and law of the iterated logarithm in the 
 case of  C-system MIXMAX generator of pseudorandom numbers.  The C-system nature of the MIXMAX generator provides well define mathematical background and guaranty  the uniformity of generated sequences.

\section{\it Classical Limit Theorems in Probability Theory}

Consider an infinite sequence of {\it independent and identically distributed random variables}  $\xi_1(x), \xi_2(x),... $ on 
the interval $0 \leq x \leq 1$ having finite mean values $\bf{M}$$\xi_{k} =\mu $ and finite variance $\sigma^2 = \bf{M}$$(\xi_k - \mu)^2$~  ( $0 < \sigma^2  < \infty $) \cite{kolmogorovprob}.  One of the fundamental  questions of interest 
in probability theory is the limiting behaviour of the sum
\cite{kolmogorovprob,khinchine,kolmo2,hartman,gendenkokolmogorov,prochorov,petrov} 
\be\label{thesum0}
S_n = \xi_1 + \xi_2+....+\xi_n = \sum^n_{k=1} \xi_k
\ee
as $n \rightarrow \infty$.  By the classical {\it central limit theorem} the difference between the average 
$S_n /  n$ and $\mu$ multiplied by the factor $\sqrt{n}$  {\it converges  in probability} 
to the normal distribution $\Phi({z\over \sigma})$ 
\be\label{clt1} 
{\bf{P}} ~\left \{ ~\sqrt{n}({S_n \over   n} - \mu) < z ~ \right \}~~ \rightarrow~~ {1 \over \sqrt{2 \pi \sigma^2}}\int^{z}_{-\infty} e^{-{y^2 \over 2 \sigma^2}} ~dy~~~~~~~~~ \text{for every z}.
\ee
The estimates of the convergence rate in the above central limit theorem were obtained 
by Lyapunov, Berry, Esseen  and others \cite{liapounoff,liapounoff1,berry,esseen,esseen1}. For independent and identically distributed random variables having finite absolute third moments $\chi^3 ={\bf M} \xi^3_k < \infty$ it has the form: 
\be
\sup_{z} \vert    {\bf{P}} ~\left \{ ~\sqrt{n}({S_n \over   n} - \mu) < z ~ \right \} - 
\Phi({z\over \sigma})  \vert \leq    {1\over   \sqrt{n}}  \Big({\chi \over \sigma }\Big)^3~.
\ee
By the Kolmogorov {\it strong law of large numbers} the average $S_n /  n$
converges {\it almost surely} to the common mean value $\mu$ of the random variables $\xi_k(x)$, that is  
\be\label{lawlargen}
{\bf{P}}   ~\left \{~  \lim_{n \rightarrow \infty}  {S_n \over   n} = \mu  ~\right \}~~ =1~.
\ee
Under the same conditions as in the above theorems Petrov \cite{petrov} has  derived the estimates of the order of growth of the sums  $S_n$ (\ref{thesum0}). The following growth estimates take place 
\bea\label{estimate}
{\bf{P}}   ~\left \{~  \lim_{n \rightarrow \infty}  { S_n - \mu \, n  \over   n^{{1\over 2} +\epsilon } } =0 ~\right \}~~ =1~,\\
{\bf{P}}   ~\left \{~  \lim_{n \rightarrow \infty}  { S_n - \mu \, n  \over   n^{{1\over 2}  }  (\ln{n})^{{1\over 2} +\epsilon }} =0 ~\right \}~~ =1~,\nn\\
{\bf{P}}   ~\left \{~  \lim_{n \rightarrow \infty}  { S_n - \mu \, n  \over   n^{{1\over 2}  }  \ln{n}^{{1\over 2} } (\ln\ln{n})^{{1\over 2} +\epsilon }} =0 ~\right \}~~ =1~,\nn\\
.........................\nn
\eea
for arbitrary $\epsilon > 0$. This result means that the random variables $S_n - \mu \, n$
cannot grow faster than $n^{{1\over 2} +\epsilon } $  or $n^{{1\over 2}  }  (\ln{n})^{{1\over 2} +\epsilon }$  or $\ln{n}^{{1\over 2} } (\ln\ln{n})^{{1\over 2} +\epsilon }$ and so on.
The theorem on the {\it law of the iterated logarithm} for a sequence of random variables 
$\left \{\xi_k\right \}$ involve conditions under which the sequence $ \lim_{n \rightarrow \infty}  \sup   {S_n - \mu \, n \over \sqrt{2 \, n  \ln \ln n }} =\sigma ~$  converges almost surely.    This relation strengthens the estimates provided by the strong law of large numbers (\ref{lawlargen}) and (\ref{estimate}).
For the independent and identically distributed random variables (\ref{thesum0})  the following Kolmogorov  {\it theorem of the iterated logarithm}  take place  \cite{khinchine,kolmo2,hartman}   
\be\label{lll}
{\bf{P}} ~\left \{ \lim_{n \rightarrow \infty}  \sup   {S_n - \mu \, n \over \sqrt{2 \, n  \ln \ln n }} = \sigma~\right \}=1,~~~~~{\bf{P}} ~\left \{ \lim_{n \rightarrow \infty}  \inf   {S_n - \mu \, n \over \sqrt{2 \, n   \ln \ln n }} = - \sigma ~\right \}=1~ ,
\ee
that is a maximal possible growth of the sum is  $\sigma \sqrt{2 \, n   \ln \ln n }$.
In order to gain an intuitive understanding of this result it is worth to calculate the probability of large  fluctuations of the sum $S_n$ using the central limiting theorem (\ref{clt1} ).  It follows from (\ref{clt1})  that for 
the arbitrary positive number $b$,~ $z=~\sigma ~ b  \sqrt{2 \ln \ln n } $ and large $n$ take place the following  relation 
\be 
{\bf{P}}~\left \{ ~\sqrt{n}({S_n \over   n} - \mu)   \geq    ~ \sigma  ~b  \sqrt{2 \ln \ln n }   ~\right \}~  \rightarrow~1- {1 \over \sqrt{2 \pi \sigma^2}}\int^{\sigma ~b \sqrt{2 \ln \ln n }}_{-\infty} e^{-{y^2 \over 2 \sigma^2}} dy~,
\ee
where the right hand side has the asymptotic  $1/ 2 b (\pi \ln \ln n)^{1/2} (\ln n)^{b^2}$ 
and  therefore 
\be\label{ineq}  
{1 \over  (\ln n )^{(1+\delta) b^2}} ~\leq ~
  {\bf{P}}~\left \{  {S_n(x) - \mu \, n \over \sqrt{2 \, n  \ln \ln n }}   \geq    ~ b~ \sigma    \right \}~  \leq ~{1 \over  (\ln n )^{b^2}}~.
\ee
Considering  the subsequence  $n = q^m$, where q is a fixed integer, one can derive 
from (\ref{ineq}) the celebrated law of the iterated logarithm  (\ref{lll}) \cite{khinchine,kolmo2,hartman,gendenkokolmogorov,petrov,prochorov}.  
The law of the iterated logarithm is a refinement of the law of large numbers (\ref{lawlargen}) and  specifies 
the global behaviour of the asymptotic sequence of the sum $S_n$ since the quantity in the limit in (\ref{lll}) depends not only on single $n$ but the totality of the remainder of the sum.
Using the central limiting theorem (\ref{clt1}) now for the fluctuations in the interval $ (-\epsilon \sqrt{2  \ln \ln n } , +\epsilon \sqrt{2  \ln \ln n } )$  one can get that 
\be
{\bf{P}}   ~\left \{ ~ \vert {S_n(x) - \mu \, n \over \sqrt{2 \, n  \ln \ln n }~} ~\vert   <   \epsilon ~\right \}~~ \rightarrow~~ {1 \over \sqrt{2 \pi \sigma^2}}\int^{\sigma ~\epsilon \sqrt{2 \ln \ln n }}_{-\sigma ~\epsilon \sqrt{2 \ln \ln n }} e^{-{y^2 \over 2 \sigma^2}} dy ~~\rightarrow~~ 1,
\ee
meaning  that the  sum (\ref{thesum0}) scaled by the factor $\sqrt{2 \, n  \ln \ln n }$  is less than any $\epsilon >0$ with probability approaching one, but will be occasionally visiting
 points in the interval $(-\sigma, \sigma)$ in accordance with the theorem (\ref{lll}).
 
Our goal is to compare the asymptotic behaviour of the sum $S_n$  which 
have been establish in the above limit  theorems in probability theory 
with the asymptotic behaviour  of the corresponding quantities defined for deterministic dynamical C-systems and specifically for the C-system which have been implemented 
into  the MIXMAX  generator \cite{konstantin,Savvidy:2015jva,Savvidy:2017mew,Savvidy:2015ida,yer1986a}.

\section{\it Classical Limit Theorems  and  Deterministic C-systems}

\begin{figure}\label{fig1}
\centering
\begin{subfigure}{0.5\textwidth}
  \centering
\includegraphics[width=7.7cm]{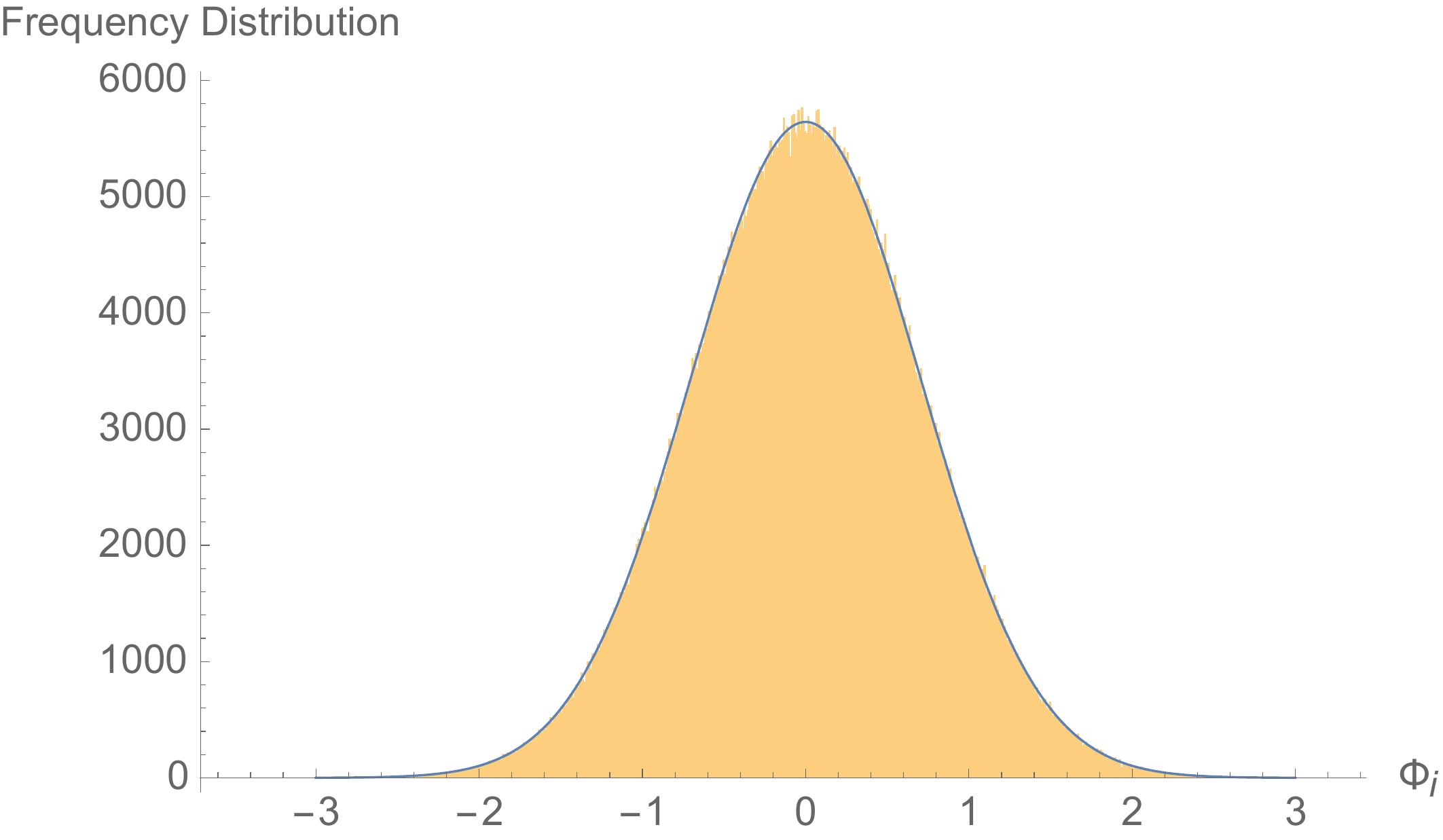}
  \caption{}
  \label{fig:sub1}
\end{subfigure}%
\begin{subfigure}{0.5\textwidth}
  \centering
  \includegraphics[width=7cm]{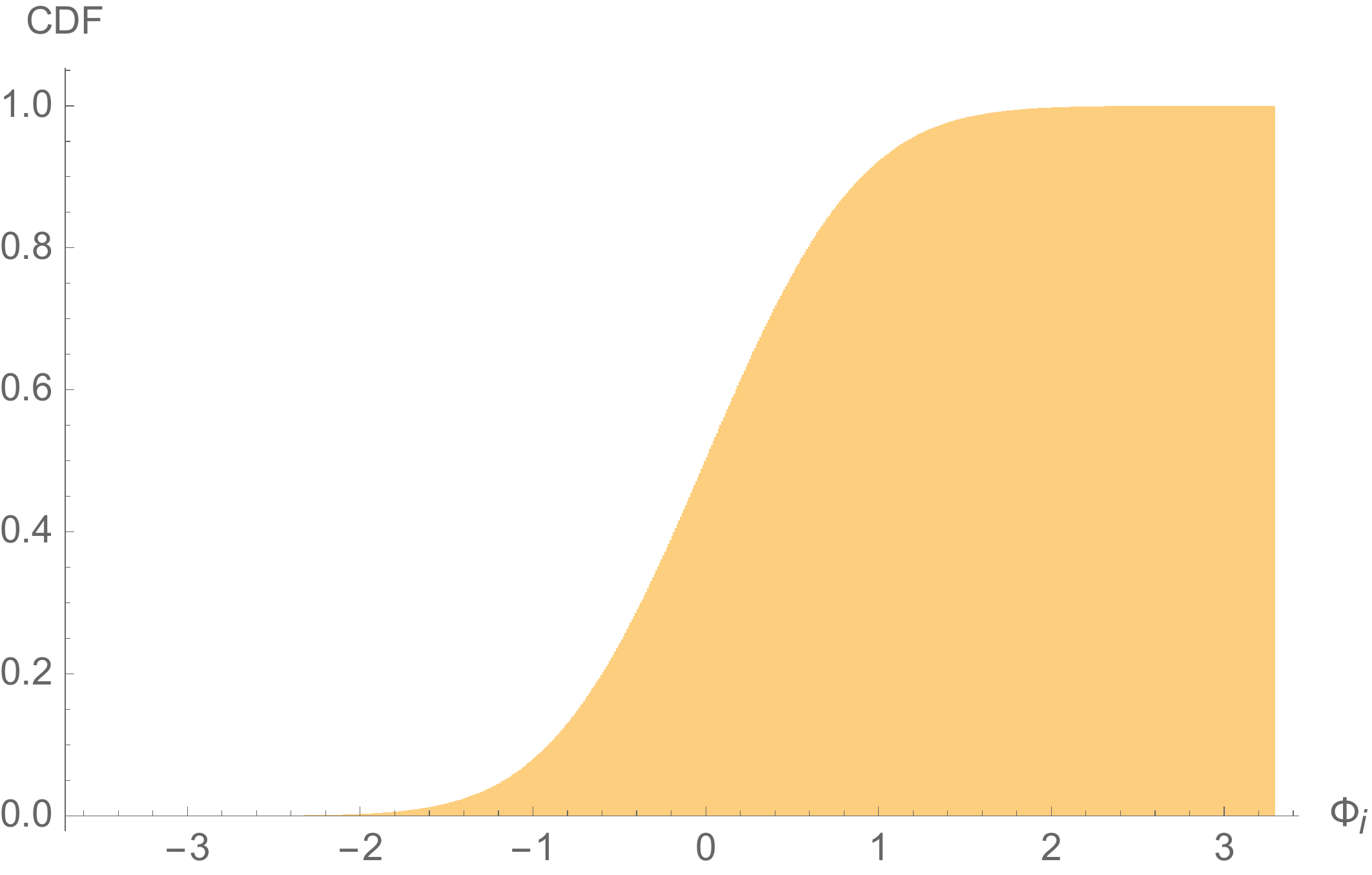}
  \caption{}
  \label{fig:sub2}
\end{subfigure}
\caption{
a)The frequency 
distribution histogram for the underlying variable $\phi_i(f,n)$ in (\ref{anderl}). The dimension of the 
C-system generator is $N=17$, the iteration time is  $n=10^4$, the 
bins are of equal size $\epsilon =0.01$ and the total number of phase space points $x_i$ is $I=10^6$.  The function $f(x) =  \cos 2 \pi (x_1+...+x_{17})$. The mean value is $\langle \phi \rangle= 0.000470253$ and the 
standard deviation $\langle \phi^2 \rangle= \sigma^2_f = 0.499867$. b) The p-value of the cumulative distribution function (CDF) for the Kolmogorov-Smirnov test is $p=0.909337$.}
\label{fig1}
\end{figure}

With that aim  let us now consider the statistical properties of deterministic dynamical C-systems. 
The hyperbolic  Anosov  C-systems defined on a torus  have {\it mixing of all orders and nonzero Kolmogorov entropy} \cite{anosov,kolmo,kolmo1,sinai3,rokhlin,rokhlin2,sinai4,gines,bowen0,ruelle,krilov,Savvidy:2017mew,yer1986a}.
The  statistical properties of a  C-system defined by the map 
 $\left \{ \forall~ x \in \CM: x \rightarrow x_n=T^n x \right \}$ 
 are characterised by  the behaviour of the  correlation functions  of observables $\left \{f(x)\right \}$ on the phase space $\CM$
   \be
\label{correlations0}
D_n(f,g)= \langle f(x) g(T^n x) \rangle - \langle f(x)\rangle \langle g( x) \rangle ,
\ee      
where $T$ denotes the Anosov C-system and $ \langle .... \rangle$ the phase space averages \cite{krilov,yer1986a,body,Gibbons:2015qja,yer1986a,body,konstantin,Savvidy:2015ida,Savvidy:2015jva}. These  correlation functions  decay exponentially, meaning that  
the observables on the phase space become independent and uncorrelated exponentially fast  \cite{Collet,moore,chernov1,young,simic,chernov,dolgopyat,tsujiit3,tsujiit2,tsujiit1,tsujiit,faure,faure1}.   For the C-systems defined on the $N$-dimensional torus  (\ref{eq:rec})
the upper bound on the exponential decay of the correlation functions 
is universal and is defined  by the value of the system entropy $h(T)$ \cite{Savvidy:2017mew} :  
\be\label{main}
\vert D_n(f,g) \vert \leq  C~  e^{-n h(T) \nu },
\ee
where $C =C(f,g)$ and $\nu=\nu(f,g)$ depend only on the observables and are 
positive numbers.    This result  allows to define the {\it decorrelation time }
$\tau_0$ for the observable $f(x)$ as  \cite{Savvidy:2017mew}
\be
\tau_0 = {1\over h(T) \nu_f}~,
\ee
where $\nu_f = \nu(f,f)$.
The index $\nu_f$ is increasing linearly $\nu_f = 2p N$ with the dimension $N$ of the C-system, where $p$ is the order 
of smoothness of the function $f(x)$ \cite{Savvidy:2017mew}. The entropy $h(T)$ 
is also increases linearly  $h(T)= {2 \over \pi} N$ (\ref{linear}) 
\cite{Savvidy:2015jva}, therefore \cite{Savvidy:2017mew}
\be
\tau_0 = {\pi \over 4 p N^2}~~.
\ee
 This result  justifies  the statistical/probabilistic description of the C-systems \cite{krilov} and 
have important consequences in the form of the law of large numbers and central limit
theorem for Anosov C-systems \cite{leonov,chernov3,rozanov,rather}.   The time average  of the observable  $f(x)$ on $\CM$ 
 \be\label{timeav}
\bar{f}_n(x) = {1 \over n} \sum^{n-1}_{k=0} f(T^k x) 
\ee
behaves  as a superposition of quantities which are statistically independent,
therefore \cite{anosov,leonov}
\be\label{avarage}
\lim_{n \rightarrow \infty} \bar{f}_n(x) = \langle f \rangle  
\ee
and the fluctuations of the time averages (\ref{timeav}) from the phase space average 
$\langle f \rangle$   multiplied by $\sqrt{n}$ 
have at large $n \rightarrow \infty$ the Gaussian distribution \cite{mac,leonov,chernov3,rozanov,rather}:
 \be\label{gauss}
\lim_{n \rightarrow \infty} m \bigg \{ x	:\sqrt{n}   \Bigg( \bar{f}_n(x)  - \langle f \rangle  \Bigg) < z    \bigg \}
= {1 \over \sqrt{2 \pi \sigma^2_f}}\int^{z}_{-\infty} e^{-{y^2 \over 2 \sigma^2_f}} dy~,
\ee
where $m$ is the invariant measure on the phase space $\CM$ and the value of the standard deviation  $\sigma_f$ is a sum
\be
\sigma^2_f = \langle f^2(x)  \rangle-
\langle f(x)  \rangle^2  +  2 \sum^{+ \infty}_{n=1} [\langle f(T^n x) f(x)  \rangle-
\langle f(x)  \rangle^2 ].  
\ee
These results allow to trace a parallel between the classical limit theorems of probability theory  and behaviour of deterministic dynamical C-systems. The theorems  (\ref{avarage}) and (\ref{gauss}) 
which taking place for the deterministic C-systems are in fine analogy with the  theorems  (\ref{lawlargen}) and (\ref{clt1}) in probability theory. This analogy can be made explicit if one use 
the  dictionary: 
\bea
\xi_k(x) &\Longleftrightarrow& f(T^k x),\nn\\
S_n(x) &\Longleftrightarrow&  \sum^{n-1}_{k=0} f(T^k x) ,\nn\\
{S_n(x) \over n} &\Longleftrightarrow&  {1\over n}\sum^{n-1}_{k=0} f(T^k x).
\eea
In the next section we shall consider the fine examples  of the C-systems defined 
on the tori. These systems represent a large class of C-systems which can be easily realised on a computer platform in the form of computer algorithms. These algorithms 
are used to generate pseudorandom numbers of high quality  and represent the called MIXMAX pseudorandom  number generators \cite{yer1986a,konstantin,Savvidy:2015ida,hepforge,root,clhep}. In particular it passes all empirical {\it U01 tests}  \cite{pierr}.

 \section{\it MIXMAX C-systems Generator }
\begin{figure}\label{fig2}
\centering
 \includegraphics[width=15cm]{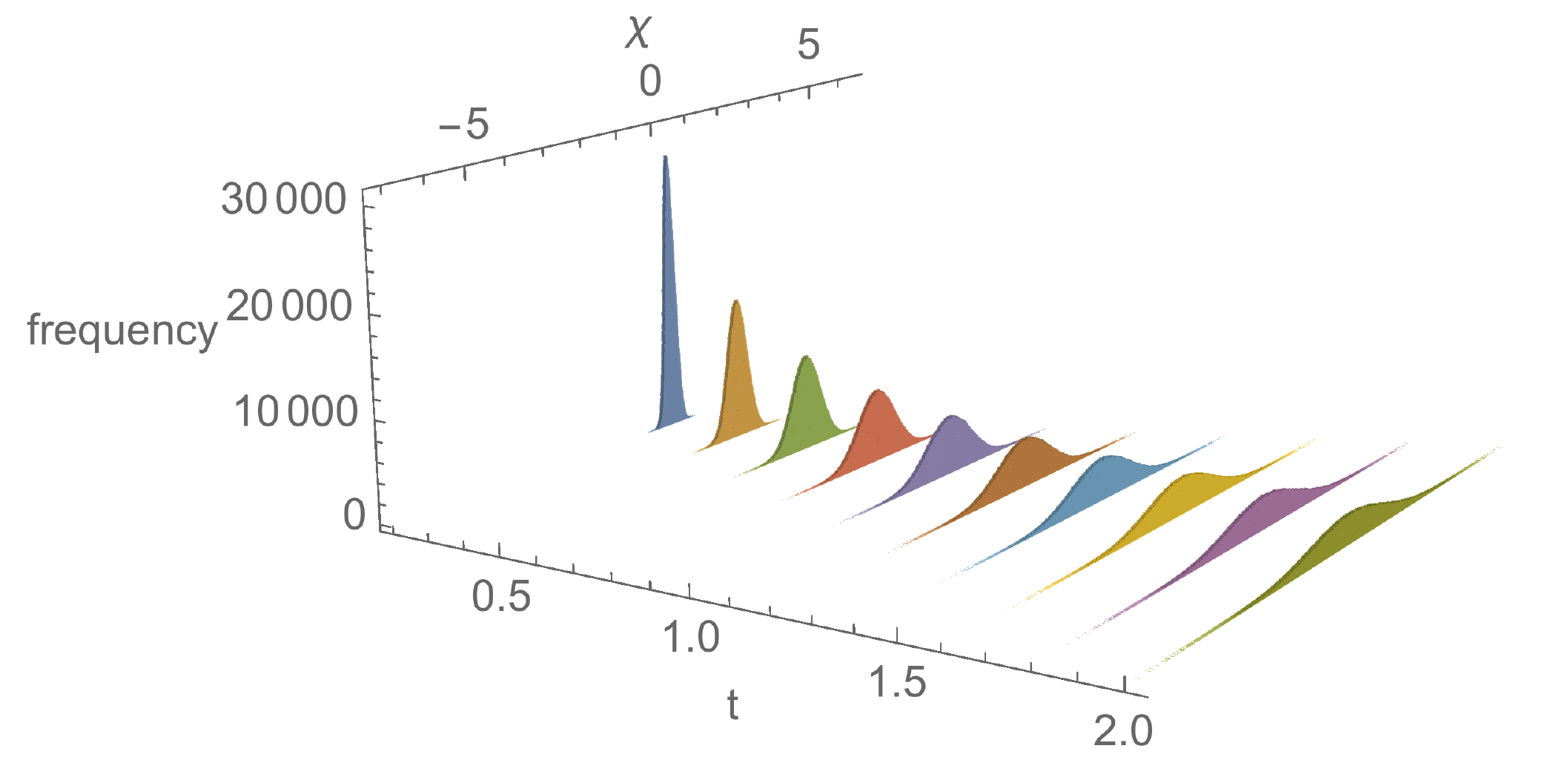}
\caption{
The frequency 
distribution histogram for the variable $\chi_i(t)$ defined in (\ref{feynman}). The "time"  interval is taken as $[0.2, 2]$. The other parameters are the same as in Fig. \ref{fig1}.}
\label{fig2}
\end{figure}

The linear automorphisms of the unit hypercube $\CM^N$ in Euclidean space $E^N$ with coordinates $ (x_1,...,x_N)$\cite{anosov,yer1986a,konstantin,Savvidy:2015ida} is defined 
as follows:
\be
\label{eq:rec}
x_i^{(k+1)} = \sum_{j=1}^N T_{ij} \, x_j^{(k)} ~~~~~\textrm{mod}~ 1,~~~~~~~~~k=0,1,2,...
\ee
where the components of the vector $x^{(k)}$ are
$
x^{(k)}= (x^{(k)}_1,...,x^{(k)}_N).
$
The phase space $\CM^N$ of the systems   (\ref{eq:rec} ) can also be 
considered as the $N$-dimensional torus \cite{anosov,yer1986a,konstantin,Savvidy:2015ida}. The operator $T$ acts on the 
initial vector $x^{(0)}$ and produces a phase space trajectory $x^{(n)}=  T^n x^{(0)}$
on a torus. The C-system is defined by the integer matrix $T$ which
has a determinant equal to one $\Det T =1$ and has no eigenvalues on the unit circle
\cite{anosov}:
\bea\label{mmatrix}
1)~\Det  T=  {\lambda_1}\,{\lambda_2}...{\lambda_N}=1,~~~~~
2)~~\vert {\lambda_i} \vert \neq 1, ~~~\forall ~~i .~~~~~~
\eea
The   measure $d m = dx_1...dx_N$ is invariant under the action of $T$.
The conditions (\ref{mmatrix}) guarantee that  $T$ represents Anosov C-system \cite{anosov} and therefore as such it is a  Kolmogorov K-system \cite{kolmo,kolmo1,sinai3,rokhlin,rokhlin2}
with mixing  of  all orders and of nonzero entropy.
 The C-system (\ref{eq:rec})  has a nonzero Kolmogorov entropy $h(T)$  \cite{anosov,sinai3,rokhlin2,sinai4,gines,Savvidy:2015ida}:
\be\label{entropyofA}
h(A) = \sum_{\vert \lambda_{\beta} \vert >1   } \ln \vert \lambda_{\beta} \vert .
\ee
We shall consider a family of matrix operators $T$ of dimension  $N$
introduced in \cite{konstantin}.  The operators $T$  fulfil the C-condition (\ref{mmatrix}) and represents a C-system \cite{yer1986a,konstantin,Savvidy:2015jva} with entropy:
\be\label{linear}
h(A)= \sum_{\beta   } \ln \vert { \lambda_{\beta} }\vert \approx
{2\over \pi}~ N
\ee
which is increases linearly with the dimension $N$ of the matrix.
Our aim is to study the asymptotic behaviour of the sum $S_n$ as $n \rightarrow \infty$ for  the pseudorandom number generator 
MIXMAX \cite{konstantin,Savvidy:2015jva} which is defined by the equations  (\ref{eq:rec}). 
\begin{figure}\label{fig:N17n500}
\centering
\begin{subfigure}{0.5\textwidth}
  \centering
  \includegraphics[width=1\linewidth]{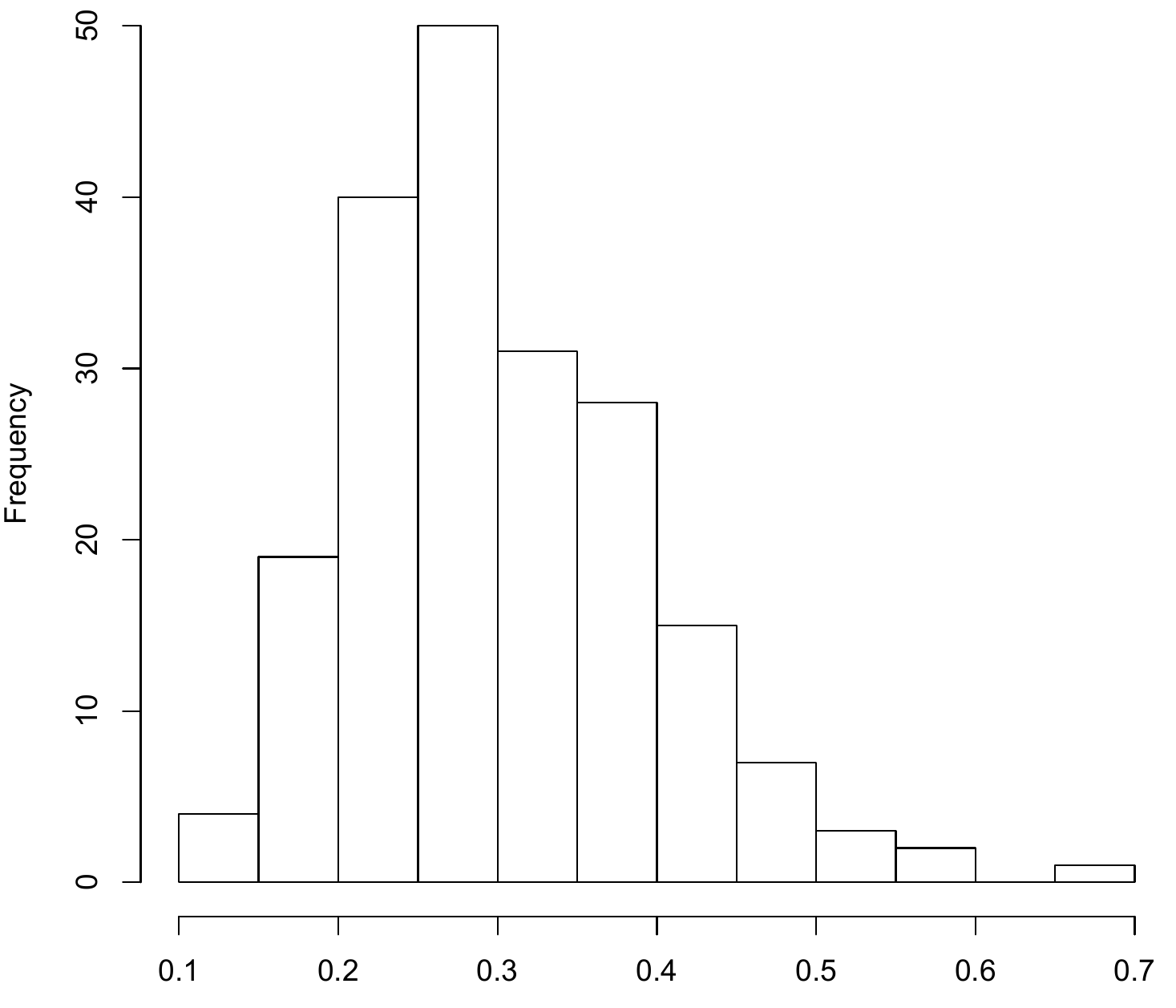}
  \label{fig:sub1}
\end{subfigure}%
\begin{subfigure}{0.5\textwidth}
  \centering
  \includegraphics[width=1\linewidth]{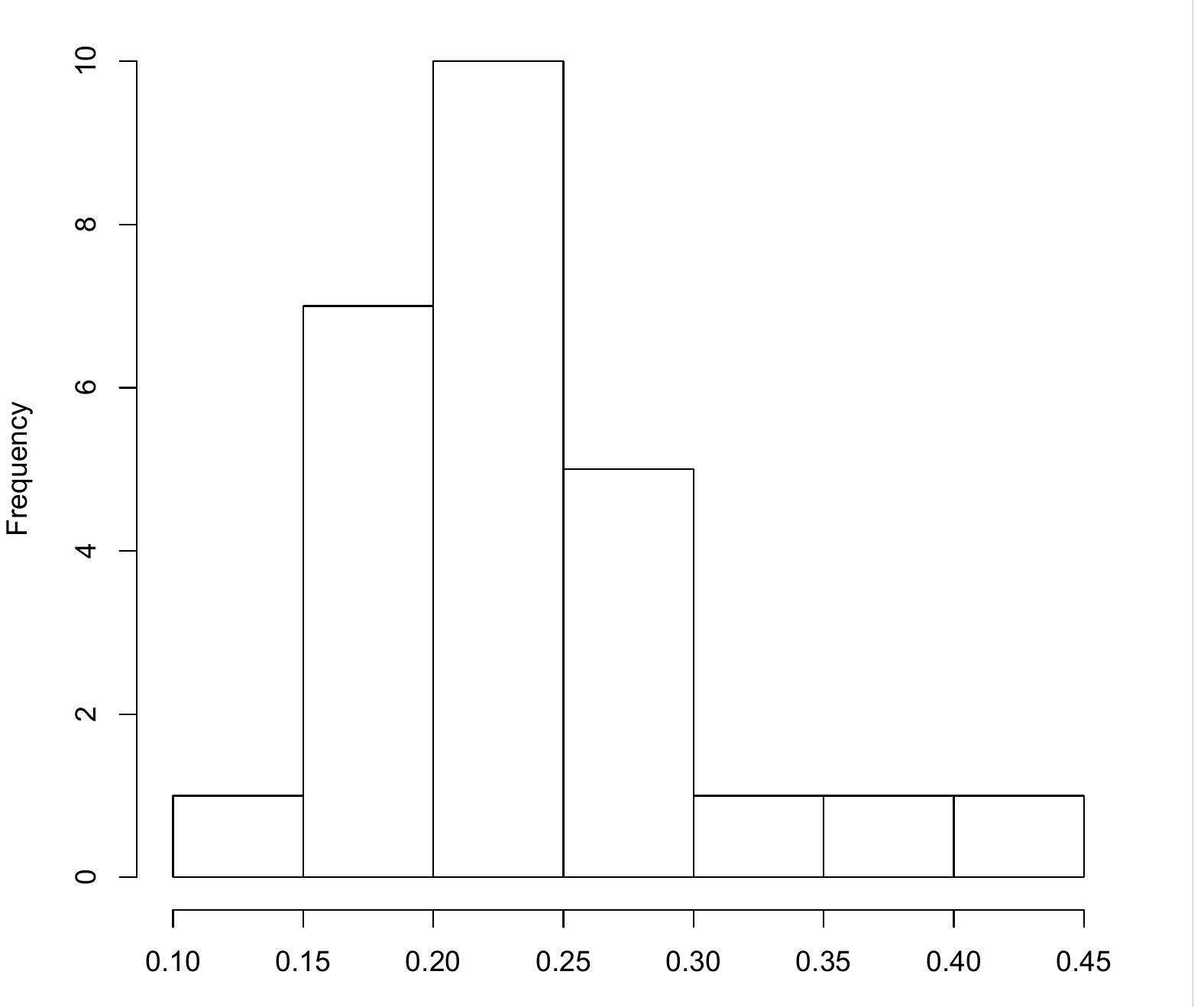}
  \label{fig:sub2}
\end{subfigure}
\caption{Two histograms of the variable $\Gamma_n=S_n/\sqrt{2 \, n \ln \ln n}$ in (\ref{kappa}).  
The dimension of the 
C-system generator is $N=240$ and  the iteration time $n =  10^9$. The total number of phase space points $x_i$ is $I=200$. Here in $S_n$ data we have subtracted the term $\mu \, n $. 
The sum $S_n$ grows approximately  as $  \sigma \sqrt{n}$.  The standard deviation 
for the observable  $f(x) = x $ is equal to $\sigma =1/\sqrt{12}  $.
The distribution function of the supremum of the  $\Gamma_n$ is calculated for the values  $n$ in the interval $n \in [m,10^9]$ for  $ m=15$ and $m=10^5$. As one can see the distribution 
of  $\Gamma_n$ for the supremum values (\ref{supremumdef}) is tightens towards $\sigma = 1/\sqrt{12} \approx 0.27$ in accordance with the Kolmogorov law of the iterated logarithm (\ref{lll}). On the first histogram,  at $m=15$, there are 65 events smaller and 85 events larger than $\sigma$. 
On the second histogram, at $m=10^5$, there are 8 events smaller and 8 events larger than $\sigma$. }  

\label{fig:3Dhis}
\end{figure}

In order to study the asymptotic  behaviour  in equation (\ref{gauss}) as $n \rightarrow \infty$ we shall consider first the following variable 
\be\label{anderl}
\phi_i(f,n)= \sqrt{n}\Big( {1 \over n} \sum^{n-1}_{k=0} f(T^k x_i)  - \langle f \rangle  \Big)
\ee
which depends on initial phase space vector  $x_i$ of the $N$-dimensional unit hypercube $ \CM^{N}$, the function $f(x)$ and the number of 
 iterations $n$.  In order to calculate the number of vectors $x_i, i=1,...,I$  which
fulfil the inequality $\phi_i(f,n) < z$  we shall construct the frequency 
distribution of the underlying variable $\phi_i(f,n)$. The bins will be taken 
of equal size $\epsilon$.
The Fig. \ref{fig1} represents the distribution function  calculated for the MIXMAX 
 generator of size $N=17$ and the 
comparison  with the Gaussian distribution
\be 
\rho(\phi) = {I \epsilon \over \sqrt{2 \pi \sigma^2_f}} ~e^{-{\phi^2 \over 2 \sigma^2_f}},
\ee
shown on the Fig.\ref{fig1} as a solid blue line. We have used the Kolmogorov-Smirnov test to calculate the $p$-value.  The p-value of the cumulative distribution function (CDF) for the Kolmogorov-Smirnov test here was $p=0.909337$. 
The null hypothesis that the data is distributed according to the normal distribution is not rejected at the $0.1 \%$   level  
based on the  Kolmogorov-Smirnov test.

Introducing a new  parameter $t = p/n$, where $p$ is an integer number $p \in \CZ$ and the alternative variable 
\be\label{feynman}
\chi_i(t) = t \sqrt{n} \Big(  {1 \over t n } \sum^{t n-1}_{k=0} f(T^k x_i)  - \langle f \rangle  \Big)
\ee
we can find the distribution function for the variable $\chi$. It was proven  that  the variable $\chi$ is described in accordance with the Wiener-Feynman process \cite{chernov1}
\be\label{diffusion}
\rho(\chi,t) = {I \epsilon \over \sqrt{2 \pi \sigma^2_f t }} ~e^{-{\chi ^2\over 2 \sigma^2_f t}}.
\ee
On Fig. \ref{fig2} one can see that with the increasing "time" $t$ the distribution 
evolve as in (\ref{diffusion}).
The useful analogy will be if one consider $I \epsilon $ as the number of "particle" at the initial time of the diffusion and $D=\sigma^2_f $ as the diffusion coefficient.

Considering the central limit theorem we were performing  
iterations for the relatively small values of $n \approx 10^4$.
In order to study the large fluctuations of the sum $S_n$ described by 
the law of the iterated logarithm we generated  sequences of increasing 
length $n=10^7 - 10^9$ and then constructed  the distribution function of the maximum values of the variable   
\be\label{kappa}
\Gamma_n(x)= {\sum^{n-1}_{k=0} f(T^k x) - \langle f \rangle ~n \over   \sqrt{2 \, n   \ln \ln n  }  }   
\ee
at the tail of the sequences in accordance with the definition of the limit superior
\be\label{supremumdef}
\overline{\lim}  ~\Gamma := 
\lim_{m \rightarrow \infty}  (~\sup_{n \geq m} \Gamma_n ~) .
\ee
We illustrated the distribution function for the values  $n$ in the interval $n \in [m,10^9]$ for  $ m=15$ and $m=10^5$ on Fig. \ref{fig:3Dhis}. As one can see the distribution 
of the supremum values is tightens towards $\sigma = 1/\sqrt{12}$.

\section{\it Acknowledgement }
This work was supported in part by the European Union's Horizon 2020
research and innovation programme under the Marie Sk\'lodowska-Curie
Grant Agreement No 644121.

\vfill


\begin{thebibliography}{99}


 \bibitem{kolmogorovprob}
A. Kolmogorov,  \emph{Grundbegriffe der Wahrscheinlichkeitsrechnung}, Berlin (1933), p. 150.


\bibitem{khinchine}
 A. Khinchine, \emph{\"Uber einen Satz der Wahrscheinlichkeitsrechnung}, 
 Fundamenta Mathematicae, {\bf 6} (1924) 9-20


 \bibitem{kolmo2} A.N. Kolmogorov,
 \emph{\"Uber das Gesetz des iterierten Logarithmus},
 Mathematische Annalen,  {\bf  101} (1929) 126-135.

 \bibitem{hartman} 
P. Hartman and A. Wintner,  
 \emph{On the law of the iterated logarithm}, Amer. J. Math.  {\bf 63} (1941) 169-176.

 \bibitem{gendenkokolmogorov}
B. V. Gnedenko and A. N. Kolmogorov, Limit distributions for sums of independent random variables . Gos.Izdat.Tech.-Theo.Lit., Moscow, 1949,  St.Petersburg; \\
  Bull. Amer. Math. Soc. {\bf 62} (1956)  50-52. 


\bibitem{petrov} V.V.Petrov, \emph{Sums of Independent Random Variables}, Izdat. "Nauka" Glav.Redak. Fiz.-Math. Lit., Moscow, 1972; https://link.springer.com/book/10.1007/978-3-642-65809-9



\bibitem{prochorov}Yu.V. Prokhorov  and V.Statulevicius, \emph{Limit theorems of probability theory}, Springer-Verlag Berlin Heidelberg New York, 2000;     Itogi Nauki i Tekhniki, Sovremennye Problemy Matematiki, Fundamental'nye Napravleniya, Vol. 81, Teoriya Veroyatnostej 6. Publisher VINITI, Moscow 1991. 


\bibitem{liapounoff} A. Liapounoff,  \emph{Sur une proposition de la th\'erie des 
probabilit\'es},  Bull. Acad. Sci. St-P\'etersbourg,  {\bf 5}  (1900) 13.

\bibitem{liapounoff1} A. Liapounoff, \emph{ Nouvelle forme du th\'eor\'eme sur la limite de probabilit\'e},  M\'em. Acad. Sci. St-P\'etersbourg, {\bf 8}  (1901) 12.



\bibitem{berry} A.C. Berry, \emph{The accuracy of the Gaussian approximation to the sum of independent variates},  Trans. Amer. Math. Soc. {\bf 49}  (1941) 122-136.  


\bibitem{esseen} C.G. Esseen,  \emph{ On the Liapounoff limit of error in the theory of probability},  Ark. Mat. Astron. Fys. A28 (1942)  1-19.

\bibitem{esseen1} C.G. Esseen,  \emph{ Fourier analysis of distribution functions}, Acta Math. 77 (1945) 1-125. 


\bibitem{petrov} V.V.  Petrov, \emph{  On the strong law of large numbers}, Teor. Veroyatn. Primen. {\bf 14} (1969) 193-202;  Theor. Probab. Appl. {\bf 14} (1969) 183-192.

\bibitem{anosov}  D. V. Anosov, \emph{Geodesic flows on closed Riemannian manifolds with negative curvature},  Trudy Mat. Inst. Steklov., Vol. {\bf 90} (1967) 3 - 210

\bibitem{kolmo} A.N. Kolmogorov,  \emph{New metrical invariant of transitive dynamical
systems and automorphisms of Lebesgue spaces},
Dokl. Acad. Nauk SSSR,  {\bf{119}} (1958) 861-865

\bibitem{kolmo1} A.N. Kolmogorov,  \emph{On the entropy per unit time as a metrical invariant
of automorphism},
Dokl. Acad. Nauk SSSR,  {\bf{124}} (1959) 754-755

\bibitem{sinai3} Ya.G. Sinai,  \emph{On the Notion of Entropy of a Dynamical System}, Doklady of Russian Academy of Sciences, {\bf{124}}  (1959)  768-771.

\bibitem{rokhlin} V.A. Rokhlin, \emph{On the endomorphisms of compact commutative groups},
Izv. Akad. Nauk, vol.  {\bf{13}} (1949) 329

\bibitem{rokhlin2}V.A. Rokhlin, \emph{On the entropy of automorphisms of compact commutative groups},
 Teor. Ver. i Pril., vol. {\bf3}, issue 3  (1961)  351


\bibitem{sinai4}Ya. G. Sinai, \emph{Proceedings of the International Congress
of Mathematicians}, Uppsala (1963) 540-559.

\bibitem{gines}A.~L.~Gines, \emph{Metrical properties of the endomorphisms on m-dimensional
torus},  Dokl. Acad. Nauk SSSR,  {\bf 138} (1961) 991-993

\bibitem{bowen0} R.Bowen, \emph{Equilibrium States and the Ergodic Theory of Anosov Diffeomorphisms}. (Lecture Notes in Mathematics, no. 470: A. Dold and B. Eckmann, editors). Springer-Verlag (Heidelberg, 1975), 108 pp.
 
\bibitem{ruelle} D. Ruelle, \emph{ Thermodynamic Formalism}, Addison-Wesley, Reading, Mass., 1978

\bibitem{krilov}N.S.Krylov, \emph{Works on the foundation of statistical physics}, M.- L. Izdatelstvo
Acad.Nauk. SSSR, 1950; (Princeton University Press, 1979)



\bibitem{yer1986a}  G. Savvidy and N. Ter-Arutyunyan-Savvidy, \emph{ On the Monte Carlo simulation of physical systems}, J.Comput.Phys. {\bf 97} (1991) 566; Preprint EFI-865-16-86-YEREVAN, Jan. 1986. 13pp.

\bibitem{konstantin} K.Savvidy, \emph{The MIXMAX random number generator},
Comput.Phys.Commun. 196 (2015) 161-165. (http://dx.doi.org/10.1016/j.cpc.2015.06.003);
 arXiv:1404.5355

 
\bibitem{Savvidy:2015jva}
  K.~Savvidy and G.~Savvidy,
 \emph{Spectrum and Entropy of C-systems. MIXMAX random number generator,}
  Chaos Solitons Fractals {\bf 91} (2016) 33
  doi:10.1016/j.chaos.2016.05.003
  [arXiv:1510.06274 [math.DS]].

 
\bibitem{Savvidy:2017mew}
  G.~Savvidy and K.~Savvidy,
 \emph{Hyperbolic Anosov C-systems. Exponential Decay of Correlation Functions,}
  arXiv:1702.03574 [math-ph].

\bibitem{Savvidy:2015ida}
  G.~Savvidy,
  \emph{Anosov C-systems and random number generators,}
  Theor.\ Math.\ Phys.\  {\bf 188} (2016) 1155;
  doi:10.1134/S004057791608002X
  [arXiv:1507.06348 [hep-th]].
   

  
\bibitem{mac}M. Kac, \emph{On the Distribution of Values of Sums of the Type $\sum f(2^{k t})$},
 Annals of Mathematics, {\bf 47} ( 1946) 33-49

\bibitem{leonov}V. P. Leonov, \emph{On the central limit theorem for ergodic endomorphisms
of the compact commutative groups}, Dokl. Acad. Nauk SSSR,  {\bf 124} No: 5 (1969) 980-983

\bibitem{chernov3} N. Chernov, \emph{ Limit theorems and Markov approximations for chaotic dynamical systems},Probab. Th. Rel. Fields  {\bf 101} (1995)  321-362 (doi:10.1007$/$BF01200500)

\bibitem{rozanov}Yu. A. Rozanov, \emph{A Central Limit Theorem for Additive Random Functions},
Theory of Probability and Its Applications,   {\bf 5}  (1960) 221-223
(doi: 10.1137$/$1105022)

\bibitem{rather} Ratner, M. \emph{The central limit theorem for geodesic flows on n-dimensional manifolds of negative curvature},
Israel J. Math. {\bf 16} (1973) 181 (doi:10.1007$/$BF02757869)



\bibitem{Collet}  P. Collet, H. Epstein and G. Gallavotti, 
\emph{ Perturbations of Geodesic Flows on Surfaces of Constant Negative Curvature
and Their Mixing Properties},  Commun. Math. Phys. {\bf 95} (1984) 61-112 


\bibitem{moore}C. C. Moore, \emph{Exponential decay of correlation coefficients for geodesic flows}, Group representations, ergodic theory, operator algebras, and mathematical physics (Berkeley, Calif., 1984), Math. Sci. Res. Inst. Publ., vol. 6, Springer, New York, 1987, pp. 163 - 181.  

\bibitem{chernov1}N. I. Chernov,  \emph{Ergodic and Statistical Properties of Piecewise Linear Hyperbolic Automorphisms of the 2-Torus}, Journal of Statistical Physics,  {\bf 69 } (1992) 111

\bibitem{young}L. S. Young,  \emph{ Decay of Correlations for Certain Quadratic Maps}, 
Commun. Math. Phys. {\bf 146} (1992) 123-138

\bibitem{simic}
S. Simic' \emph{Lipschitz distributions and Anosov flows},
Proceedings of the
American Mathematical Society, {\bf 124} (1996) 1869




\bibitem{chernov}N. I. Chernov,  \emph{ Markov Approximations and Decay of Correlations for Anosov Flows},  Annals of Mathematics Second Series. {\bf 147} (1998) 269-324

\bibitem{dolgopyat}D. Dolgopyat,  \emph{ On Decay of Correlations in Anosov Flows},
Annals of Mathematics Second Series,  {\bf 147}  (1998)  357-390


\bibitem{tsujiit3}  V.Baladi and M. Tsujii, 
\emph{Anisotropic H\"older and Sobolev spaces for hyperbolic diffeomorphisms},
Annales de l'Institute Fourier, {\bf  57} (2007) 127-154;(doi:10.5802/aif.2253)

\bibitem{tsujiit2}  M. Tsujii, 
\emph{Decay of correlations in suspension semi-flows of angle-multiplying maps},
Ergod. Theo. and Dynam. Sys.,{\bf  28}  (2008) 291-317, doi: 10.1017/S0143385707000430

\bibitem{tsujiit1}  M. Tsujii, 
\emph{Quasi-compactness of Transfer Operators for Contact Anosov Flows}
arXiv:0806.0732v3 [math.DS] 8 Apr. 2010


\bibitem{tsujiit}  M. Tsujii, 
\emph{ Contact Anosov Flows and the FBI Transform}, arXiv:1010.0396v3 [math.DS] 22 Jun. 2011.


\bibitem{faure} F. Faure and J. Sj\"ostrand \emph{Upper bound on the density of Ruelle resonances for Anosov flows}, arXiv:1003.0513v1 [math-ph] 2 Mar. 2010


\bibitem{faure1}F. Faure and J. Sj\"ostrand \emph{Semi-classical approach for Anosov diffeomorphisms and Ruelle resonances}
arXiv:0802.1780v3 [nlin.CD] 8 Sep 2008


\bibitem{yangmillsmech} G.Savvidy, \emph{The Yang-Mills mechanics as a Kolmogorov K-system}, Phys.Lett.B {\bf{130}} (1983) 303
   
 
\bibitem{Savvidy:1982jk}
  G.~Savvidy,
\emph{Classical and Quantum Mechanics of Nonabelian Gauge Fields,}
  Nucl.\ Phys.\ B {\bf 246} (1984) 302.

\bibitem{body} V.Gurzadyan and G.Savvidy, \emph{Collective relaxation of stellar systems},
Astron. Astrophys. {\bf 160} (1986) 203

\bibitem{Gibbons:2015qja}
  G.~W.~Gibbons,
\emph{The Jacobi-metric for timelike geodesics in static spacetimes,}
  Class.\ Quant.\ Grav.\  {\bf 33} (2016) no.2,  025004
  doi:10.1088/0264-9381/33/2/025004
  [arXiv:1508.06755 [gr-qc]].
  
 
 
\bibitem{pierr} P. L'Ecuyer and R. Simard, \emph{TestU01: A C Library for Empirical Testing of Random Number Generators},  ACM Transactions on Mathematical Software, {\bf 33} (2007) 1-40.


\bibitem{hepforge} HEPFORGE.ORG, {http://mixmax.hepforge.org};\\
\url {http://www.inp.demokritos.gr/~savvidy/mixmax.php}
 
\bibitem{root}  ROOT, Release 6.04/06 on  2015-10-13

\bibitem{clhep} GEANT/CLHEP, Release 2.3.1.1 on 2015-11-10.

  
  
\end{thebibliography}
\end{document}